\documentclass
[pre,twocolumn,superscriptaddress,showpacs,10pt]{revtex4}%
\usepackage{amsfonts}
\usepackage{amsmath}
\usepackage{amssymb}
\usepackage{graphicx}%
\usepackage{mathptmx}

\begin{document}


\title{Stability of two-dimensional ion-acoustic wave packets in quantum plasmas}

\author{Amar P. Misra}
\email{apmisra@visva-bharati.ac.in}
\altaffiliation{Permanent address: Department of Mathematics, Siksha Bhavana, Visva-Bharati
University, Santiniketan-731 235, India}
\affiliation{Department of Physics, Ume\aa \ University, SE-901 87 Ume\aa , Sweden}

\author{Mattias Marklund}
\email{mattias.marklund@physics.umu.se}
\affiliation{Department of Physics, Ume\aa \ University, SE-901 87 Ume\aa , Sweden}

\author{Gert Brodin}
\email{gert.brodin@physics.umu.se}
\affiliation{Department of Physics, Ume\aa \ University, SE-901 87 Ume\aa , Sweden}

\author{Padma K. Shukla}
\email{ps@tp4.rub.de}
\affiliation{Institut f\"{u}r Theoretische Physik IV, Fakult\"{a}t f\"{u}r Physik and
Astronomie, Ruhr-Universit\"{a}t Bochum, D-44780 Bochum, Germany}

\date{22 Dec., 2009}

\begin{abstract}
The nonlinear propagation of two-dimensional (2D) quantum ion-acoustic waves
(QIAWs) is studied in a quantum electron-ion plasma. By using a 2D quantum
hydrodynamic model and the method of multiple scales, a new set of
coupled nonlinear partial differential equations is derived which
governs the slow modulation of the 2D QIAW packets. The oblique
modulational instability (MI) is then studied by means of a corresponding nonlinear
Schr\"{o}dinger equation derived from the coupled nonlinear partial differential equations. It is shown that the
quantum parameter $H \propto \hbar$, associated with the Bohm potential, shifts the MI
domains around the $k\theta$-plane, where $k$ is the carrier wave number and
$\theta$ is the angle of modulation. In particular, the
ion-acoustic wave (IAW), previously known to be stable under parallel
modulation in classical plasmas, is shown to be unstable in quantum
plasmas. The growth rate of the MI is found to be quenched by the obliqueness of
modulation. The modulation of 2D QIAW packets along $\mathbf{k}$ is shown to
be described by a set of Davey-Stewartson-like equations. The latter can
be studied for the 2D wave collapse in dense plasmas. The predicted results,
which could be important to look for stable wave propagation in laboratory
experiments as well as in dense astrophysical plasmas, thus generalize the
theory of MI of IAW propagations both in classical and quantum electron-ion plasmas.
\end{abstract}


\maketitle

\section{Introduction}

The importance of quantum effects has been recognized over the last few years
in view of its remarkable applications in metallic and semiconductor
nanostructures (e.g., metallic nanoparticles, metal clusters, thin metal
films, nanotubes, quantum well and quantum dots, nano-plasmonic devices etc.)
\cite{Marklund1,Markowich,Shukla1,Shukla2,Marklund2,Marklund3,Brodin1,Brodin2}
as well as in dense astrophysical environments (e.g., white dwarfs, neutron
stars, supernovae etc.) \cite{Manfredi,Silva}.

It is well known that the nonlinear propagation of wave packets in a
dispersive plasma medium is generically subject to amplitude modulations
due to the carrier wave self-interaction, i.e., a slow variation of the wave
packet's envelope due to nonlinearities. Under certain conditions, the system's
evolution may thus undergo a modulational
instability (MI), leading to energy localization via the formation of envelope
solitons. Such solitons are governed by a nonlinear Schr\"{o}dinger (NLS)
equation where the nonlinearity at the first stage of the amplitude is
balanced by the group dispersion. This mechanism is encountered in various
physical contexts, including pulse formation in nonlinear optics, in material
science, as well as in plasma physics. A number of works can be found in the
literature for the investigation of MI in classical (see, e.g., Refs.\
\cite{Kako,Watanabe,Mishra,Amin,Misra,Kourakis,Kalejahi}) as well as quantum
plasmas (see e.g., Refs.\ \cite{Misra1,Misra2,Misra3,Sabry}). The MI of
ion-acoustic waves (IAWs) has been shown to be a general property in a
nonlinear dispersive plasma medium, when the modulation is considered
obliquely to the direction of propagation of the wave vector \cite{Kako}. Experimental observations of the MI of IAWs have been reported by Watanabe
\cite{Watanabe}. Stable wave propagation from modulational obliqueness, in both
classical \cite{Mishra,Amin,Kourakis,Kalejahi} and quantum plasmas
\cite{Misra3}, has also been investigated by a number of authors. 

On the other hand, in a wide variety of
scientific fields (e.g., nonlinear optics, plasma physics, fluid dynamics
etc.), certain nonlinear governing equations often exhibit important phenomena other than solitons, such as shocks
(wave singularity), self-similar structures, wave collapse (i.e., blow up with
growing amplitude or amplitude decay at finite time or finite distance of
propagation), as well as the wave radiation emission leading to the onset of
chaos. One such important system in this context is the Davey-Stewartson (DS)-like equations
\cite{Davey}, where the system has both quadratic as well as cubic
nonlinearities. Such equations can appear not only in the field of fluid
dynamics as in the case of water wave propagation \cite{Davey,Ablowitz}, or in
optical communications and information processing in nonlinear optical media
\cite{Hasegawa,Haus,Wu}, but also in plasma physics community \cite{Xue,Gill}.
The DS description of collective excitations in Bose-Einstein condenstates
has also been studied in the context of matter-wave solitons
\cite{Huang,Skupin}.

The purpose of the present work is to investigate, in more detail, the
stability and instability criteria for the modulation of quantum IAW (QIAW)
packets using two-dimensional (2D) quantum fluid model, and to generalize the
previous investigation \cite{Kako}, both from classical and quantum points of
view. Here we show that the nonlinear dynamics of QIAWs as well as the static
zeroth harmonic field is governed by a new set of coupled nonlinear partial
differential equations, which, in particular, reduces to a set of Davey-Stewartson (DS)-like equations for surface water waves \cite{Davey}. We
show that the quantum parameter $H \propto \hbar$, associated with the Bohm potential,
shifts the MI domains in the $k\theta$-plane in 2D quantum plasmas. Here $k$
is the carrier wave number and $\theta$ is the angle of modulation with the
propagation vector. We also show that the maximum growth rate of MI can be
reduced by the obliquenes parameter $\theta$ rather than $H$ as in
one-dimensional (1D) quantum plasmas \cite{Misra1,Misra2}. Moreover, a
criterion for the existence of 2D wave collapse is presented from the DS-like equations.

The paper is organized as follows. In Sec.\ II, we describe the 2D quantum
hydrodynamic model, and derive the governing system of equations
using the method of multiple scales. In Sec.\ III the MI of the QIAWs
is studied and the condition for the existence of 2D wave collapse is
derived. Finally, the Sec.\ IV is left for concluding the results.

\section{Fluid model and derivation of the evolution equations}

We consider the nonlinear propagation of QIAWs in a 2D quantum electron-ion
plasma. The basic normalized set of equations in a two-component unmagnetized
quantum plasma reads%
\begin{eqnarray}
&&
\frac{\partial n_{i}}{\partial t}+\nabla\cdot(n_{i}\mathbf{V})=0, \label{e1}%
\\ &&
\frac{\partial\mathbf{V}}{\partial t}+(\mathbf{V}\cdot\nabla)\mathbf{V}%
=-\nabla\phi, \label{e2}%
\\ &&
\nabla\phi-\frac{1}{3}n_{e}^{-1/3}\nabla n_{e}+\frac{H^{2}}{2}\nabla\left(
\frac{\nabla^{2}\sqrt{n_{e}}}{\sqrt{n_{e}}}\right) = 0  , \label{e3}%
\\ &&
\nabla^{2}\phi=n_{e}-n_{i}, \label{e4}%
\end{eqnarray}
where $\nabla\equiv(\partial/\partial x,\partial/\partial y),$ $n_{e(i)}$ is
the electron (ion) number density normalized by their equilibrium value
$n_{0}$, $\mathbf{V\equiv(}u\mathbf{,}v\mathbf{)}$ is the ion velocity
normalized by the ion-acoustic speed $c_{s}=\sqrt{k_{B}T_{Fe}/m_{i}}$ with
$k_{B}$ denoting the Boltzmann constant, $m_{i}$ the ion mass, $T_{Fe}%
\equiv\hbar^{2}(3\pi^{2}n_{0})^{2/3}/2k_{B}m_{e}$ the electron Fermi
temperature, and $\hbar$ the scaled Planck's constant. Also, $H=\hbar
\omega_{pe}/k_{B}T_{Fe}$ is denoting the ratio of the `plasmon energy density'
to the Fermi thermal energy, where $\omega_{pj}=\sqrt{n_{0}e^{2}%
/\varepsilon_{0}m_{j}}$ is the plasma frequency for the $j$-th particle.
Moreover, $\phi$ is the electrostatic potential normalized by $k_{B}T_{Fe}%
/e.$The space and time variables are respectively normalized by $c_{s}%
/\omega_{pi}$ and the inverse of $\omega_{pi}.$ The electron pressure gradient
$(\nabla p_{e})$ and quantum force in Eq. (\ref{e3}) appear due to the
electron degeneracy in a dense plasma with the Fermi distribution function.
The former can be given (since the equilibrium pressure is truly
three-dimensional) by the following equation of state \cite{Manfredi,Landau}.
\begin{equation}
p_{e}=\frac{1}{5}\frac{m_{e}V_{Fe}^{2}}{n_{0}^{2/3}}n_{e}^{5/3} \label{e5}%
\end{equation}
where $V_{Fe}\equiv\sqrt{k_{B}T_{Fe}/m_{e}}$ is the Fermi thermal speed of electrons.

In order to obtain the evloution equations for QIAW packets, we employ the
standard multiple-scale technique (MST) \cite{Tanuity} \ in which the space
and the time variables are stretched as%
\begin{equation}
\xi=\epsilon(x-v_{gx}t),\eta=\epsilon(y-v_{gy}t),\tau=\epsilon^{2}t,
\label{e6}%
\end{equation}
where $\epsilon$ is a small parameter representing the strength of the wave
amplitude and $\mathbf{v}_{g}\equiv(v_{gx},v_{gy})$ is the normalized (by
$c_{s})$ group velocity, to be determined later by the compatibility
condition. The dynamical variables are expanded as%
\begin{eqnarray}
&&
n_{j}=1+\sum_{n=1}^{\infty}\epsilon^{n}\sum_{l=-\infty}^{\infty}n_{jl}%
^{(n)}(\xi,\eta,\tau)\exp[i(\mathbf{k\cdot r}-\omega t)l], \label{e7}%
\\ &&
(u,v)   =\sum_{n=1}^{\infty}\epsilon^{n}\sum_{l=-\infty}^{\infty}%
[u_{l}^{(n)}(\xi,\eta,\tau),v_{l}^{(n)}(\xi,\eta,\tau)]
\nonumber\\ && \qquad
\times\exp[i(\mathbf{k\cdot r}-\omega t)l], \label{e8}%
\\ &&
\phi=\sum_{n=1}^{\infty}\epsilon^{n}\sum_{l=-\infty}^{\infty}\phi_{l}%
^{(n)}(\xi,\eta,\tau)\exp[i(\mathbf{k\cdot r}-\omega t)l], \label{e9}%
\end{eqnarray}
where $n_{jl}^{(n)},$ $u_{l}^{(n)},$ $v_{l}^{(n)}$ and $\phi_{l}^{(n)}$
satisfy the reality condition $S_{-l}^{(n)}=S_{l}^{(n)\ast}$ with asterisk
denoting the complex conjugate of the corresponding quantity. Notice that in
Eqs. (\ref{e7})-(\ref{e9}) the perturbed states depend on the fast scales via
the phase $(\mathbf{k\cdot r}-\omega t)$ (where $\mathbf{k}$ and $\omega$
respectively denote the wave vector and wave frequency), whereas the slow
scales only enter the $l$-th harmonic amplitude. We suppose that $\mathbf{k}$
makes an angle $\theta$ with the $x$-axis, and the modulation is along any
direction in the $xy$-plane. In a general way, the wave vector $\mathbf{k}$ is
then $\mathbf{k}\equiv(k_{x},k_{y})=(k\cos\theta,k\sin\theta)$ and the group
velocity, $\mathbf{v}_{g}\equiv(v_{gx},v_{gy})=(v_{g}\cos\theta,v_{g}%
\sin\theta)$.

Substituting the expressions from Eqs. (\ref{e7})-(\ref{e9}) into the Eqs.
(\ref{e1})-(\ref{e4}) and collecting the terms in different powers of
$\epsilon$ we obtain for $n=1,l=1$ the linear dispersion relation for the
normalized wave frequency ($\omega \rightarrow \omega/\omega_{pi}$) and the wave number ($k \rightarrow kc_s/\omega_{pi}$) as%
\begin{equation}
\omega^{2}=\frac{k^{2}\Lambda}{1+k^{2}\Lambda}, \label{e10}%
\end{equation}
where $\Lambda\equiv1/3+H^{2}k^{2}/4.$ Note that the wave number $k$ in Eq.
(\ref{e10}) is not very small such that the wave frequency $\omega$ is
sufficiently large to prevent the appearance of harmonic modes of $k_{x}%
,k_{y}$ as proper modes. These harmonic modes will virtually appear in higher
orders of perturbations. Also, since $k$ is normalized by the inverse of Fermi
screening length ($\lambda_{F}\equiv c_{s}/\omega_{pi}$), the values of $k$
greater than unity is inadmissible, otherwise the wavelength would become
smaller than the screening length. As a result, the collective behaviors of
the plasma will disappear. Moreover, we consider the quantum parameter $H,$ to
vary in the range $0.1<H\lesssim0.45$ such that the ratio $V_{Fe}/c \ll 1$ [an
approximate condition for the nonrelativistic quantum hydrodynamic model
to be valid] and the coupling parameter, $g_{Q}\equiv2m_{e}e^{2}%
/\varepsilon_{0}\hbar^{2}(3\pi^{2}\sqrt{n_{0}})^{2/3}\lesssim1$ (which
corresponds to the density region where the quantum collective and mean field
effects become important).

On the other hand, for the second order reduced equations with $n=2,l=1$, we
obtain the following compatibility conditions for the group velocity components%
\begin{align}
v_{gx}  &  \equiv\frac{\partial\omega}{\partial k_{x}}=\frac{\omega^{3}k_{x}%
}{k^{2}}\left[  \frac{1}{\omega^{2}}+\frac{H^{2}}{4\Lambda}-1\right]
,\label{e11}\\
v_{gy}  &  \equiv\frac{\partial\omega}{\partial k_{y}}=\frac{\omega^{3}k_{y}%
}{k^{2}}\left[  \frac{1}{\omega^{2}}+\frac{H^{2}}{4\Lambda}-1\right]  .
\label{e12}%
\end{align}
Next, proceeding\ in the same way as in Refs. \cite{Misra1,Misra2,Sabry} and
finally considering the equations for $l=1$ and $n=3$, we obtain the following
coupled equations for the propagation of modulated QIAWs
\begin{eqnarray}
&&\!\!\!
i\frac{\partial\phi}{\partial\tau}+P_{1}^{q}\frac{\partial^{2}\phi}%
{\partial\xi^{2}}+P_{2}^{q}\frac{\partial^{2}\phi}{\partial\eta^{2}}+P_{3}%
^{q}\frac{\partial^{2}\phi}{\partial\xi\partial\eta}+Q_{1}^{q}|\phi|^{2}\phi
\nonumber  \\ &&
+Q_{2}^{q}\psi\phi+Q_{3}^{q}\tilde{\psi}\phi+Q_{4}^{q}\bar{\psi}\phi+Q_{5}%
^{q}\chi\phi+Q_{6}^{q}\tilde{\chi}\phi=0, \label{e13}%
\\ &&
R_{1}^{q}\frac{\partial\psi}{\partial\xi}+R_{2}^{q}\frac{\partial\bar{\psi
}}{\partial\eta}+R_{3}^{q}\frac{\partial\bar{\psi}}{\partial\xi}+R_{4}%
^{q}\frac{\partial\tilde{\psi}}{\partial\eta}
\nonumber\\ && \quad 
=S_{1}^{q}\frac{\partial|\phi|^{2}}{\partial\xi}+S_{2}^{q}\frac
{\partial|\phi|^{2}}{\partial\eta}+S_{3}^{q}\frac{\partial\chi}{\partial\xi
}+S_{4}^{q}\frac{\partial\tilde{\chi}}{\partial\eta}, \label{e14}%
\end{eqnarray}
where $\phi\equiv\phi_{1}^{(1)},\bar{\psi}\equiv v_{0}^{(2)},\psi\equiv
\int\partial_{\xi}v_{0}^{(2)}\partial\eta,\tilde{\psi}\equiv\int\partial
_{\eta}v_{0}^{(2)}\partial\xi,\chi\equiv\int\partial_{\xi}|\phi|^{2}%
\partial\eta,\tilde{\chi}\equiv\int\partial_{\eta}|\phi|^{2}\partial\xi,$ and
the coefficients $P_{j}^{q},Q_{j}^{q},R_{j}^{q},S_{j}^{q}$ are given in
Appendix A, where $j=1,...,6$ for $Q$'s, $j=1,2,3$ for $P$'s and $j=1,...,4$
for others. The corresponding coefficients $P_{j}^{c},Q_{j}^{c},R_{j}%
^{c},S_{j}^{c}$ for the propagation of 2D classical IAWs can also be obtained
by the similar method as discussed in Appendix B.\ The superscripts `$q$' and
`$c$' are used to denote the coefficients corresponding to 2D quantum and 2D
classical electron-ion plasmas. Thus, we have obtained a new system of
nonlocal nonlinear equations, which describe the slow (and general) modulation
of the QIAW packets in 2D quantum plasmas. The coefficients $P_{1}^{q}%
,P_{2}^{q}$ appear due to the wave group dispersion and 2D motion, and
$P_{3}^{q}$ for the arbitrary orientations of the carrier wave propagation as
well as the modulation of the QIAW packets. The nonlinear coefficients
$Q_{1}^{q}$ is due to the carrier wave self-interaction originating from the
zeroth harmonic modes (or slow modes), i.e., the ponderomotive force,
and $Q_{5}^{q},Q_{6}^{q}$ are for the combined effects of 2D motion and
self-interaction. The nonlinear-nonlocal coefficients $Q_{2}^{q},Q_{3}%
^{q},Q_{4}^{q}$ arise due to the coupling between the dynamical field
associated with the first harmonic (with a `cascaded' effect from the second
harmonic) and a static field generated due to the mean motion (zeroth
harmonic) in the plasma. The appearance of the coefficients $R_{j}^{q}%
,S_{j}^{q}$ can also be explained similarly.

Equations (\ref{e13}) and (\ref{e14}) can be studied, as for example, for the
modulation of Stokes wave train (plane wave) with constant amplitude to a
small perturbation as well as to look for envelope solitons, and wave
collapse, if any. There are, of course, many other physical insights (e.g.,
dromion solution), that can also be recovered from this system. In particular,
looking for the modulation of QIAW packets parallel to the carrier wave
vector, one can obtain the DS-like equations \cite{Davey} in 2D quantum plasmas%
\begin{eqnarray}
&&\!\!\!\!
i\frac{\partial\phi}{\partial\tau}+P_{1}^{q}\frac{\partial^{2}\phi}%
{\partial\xi^{2}}+P_{2}^{q}\frac{\partial^{2}\phi}{\partial\eta^{2}}%
+Q_{11}^{q}|\phi|^{2}\phi+Q_{22}^{q}\psi\phi=0, \label{15}%
\\ &&
R_{1}\frac{\partial^{2}\psi}{\partial\xi^{2}}+R_{2}\frac{\partial^{2}\psi
}{\partial\eta^{2}}=S_{1}\frac{\partial^{2}|\phi|^{2}}{\partial\xi^{2}},
\label{e16}%
\end{eqnarray}
where the coefficients are those obtained at $\theta=0$ from the general
expressions. Also, $Q_{11}^{q}=Q_{1}^{q}+Q_{3}^{q}S_{1}^{q}/R_{2}^{q}$ and
$Q_{22}^{q}=Q_{2}^{q}-Q_{3}^{q}R_{1}^{q}/R_{2}^{q}.$ Similar forms like Eqs.
(\ref{15}) and (\ref{e16}) can also be obtained by transforming the $x$-axis
through an angle $\theta,$ so that the wave vector $\mathbf{k}$ and the
modulation direction coincide. \ Now, our aim is to investigate the MI of
QIAWs by considering its modulation along the $x$-axis through an angle
$\theta$ with the carrier wave vector $\mathbf{k}$. To this end, we disregard
the group velocity component of the modulated wave along the $y$-axis, as well
as the $y$- or $\eta$- dependence of the physical variables. Thus, we obtain
from Eqs. (\ref{e13}) and (\ref{e14}) the NLS equation for the oblique
modulation of QIAW packets
\begin{equation}
i\frac{\partial\phi}{\partial\tau}+P\frac{\partial^{2}\phi}{\partial\xi^{2}%
}+Q|\phi|^{2}\phi=0, \label{e17}%
\end{equation}
where $P\equiv P_{1}^{q}$ and $Q\equiv Q_{1}^{q}+Q_{4}^{q}S_{1}^{q}/R_{3}%
^{q}.$

\section{ Analysis of modulation instability}

We consider the MI of a plane wave solution of Eq. (\ref{e17}) for $\phi$ with
constant amplitude $\phi_{0}.$ The boundary conditions that $\phi\rightarrow0$
as $\xi\rightarrow\infty$ must now be relaxed, because the plane wave train is
still unmodulated and the solution is not unique$.$ Thus, we can represent the
solution as the monochromatic solution $\phi=\phi_{0}$exp$(iQ|\phi_{0}%
|^{2}\tau)$, where $\Delta(\tau)=-Q|\phi_{0}|^{2}$ is the nonlinear frequency
shift. To study the stability of this solution we modulate the amplitude
against linear perturbation as $\phi=[\phi_{0}+\phi_{m}$cos$(K\xi-\Omega
\tau)]$exp$\left(  iQ|\phi_{0}|^{2}\tau\right)  $, where $K$ and $\Omega$ are,
respectively, the wave number and the frequency of modulation. We\ then
readily obtain from Eq. (\ref{e17}) the dispersion relation
\begin{equation}
\Omega^{2}=P^2K^4\left(  1-\frac{K_{c}^{2}(\tau)}{K^{2}}\right)  ,
\label{e18}%
\end{equation}
where $K_{c}=\sqrt{2Q/P}|\phi_{0}|$ is the critical wave number such that the
MI sets in for a wave number $K<K_{c}$, i.e. for all wavelengths above the
threshold, $\lambda_{c}=2\pi/K_{c}$. The instability growth rate (letting
$\Omega=i\Gamma)$ is then given by
\begin{equation}
\Gamma=PK^{2}\sqrt{\frac{K_{c}^{2}}{K^{2}}-1,} \label{e19}%
\end{equation}
The maximum growth rate, $\Gamma_{\text{max}}=|Q||\phi_{0}|^{2}$ as equals to
the amount of the nonlinear frequency shift $|\Delta(\tau)|$, is achieved at
$K=K_{c}/\sqrt{2}$. Clearly, the instability condition depends only on
the sign of the product $PQ$, which, in turn, depends on the angle of
modulation $\theta$ as well as the quantum parameter $H$. These may be studied
numerically, relying on the exact expressions of $P$ and $Q$. \ We find that
for a fixed value of $H=0.3$ (i.e., as one enters the density region
$n_{0}=1.5295\times10^{33}$m$^{-3}$), $P>0$ in $0.6958\lesssim\theta<\pi/2$
and $Q<0$ in $0<\theta\lesssim0.5281$ and $1.138\lesssim\theta<\pi/2$ for
$0<k<1$. Other domains of $k$ and $\theta$ where $P$ and $Q$ change sign
depend on the $P=0$ and $Q=0$ curves in the $k\theta$-plane.\ Similarly,
considering another value of $H=0.45$ (where $n_{0}=1.34\times10^{32}$m$^{-3}%
$) one can find $P>0$ in $0.6748\lesssim\theta<\pi/2$ and $Q<0$ in
$0<\theta\lesssim0.6$ and $1.137\lesssim\theta<\pi/2$ for $0<k<1.$ As above,
$P,Q$ also change their sign in other regions in the $k\theta$-plane.

Figure 1 displays the contour plots of $PQ=0$ boundary curves against the
normalized wave number $k$ and the modulation angle $\theta$. The upper panel
corresponds to the quantum case, whereas the lower one is for the classical
electron-ion plasmas. The stable ($PQ<0$) and unstable ($PQ>0$) regions are
separated in the $k\theta$-plane by the boundary curves of $PQ=0$ as
indicated by the white and the gray regions respectively. We have allowed
$k$\ to vary in a range $0<k<1$ (as explained in Sec. II) and $\theta$ to vary
between $\theta=0$ and $\theta=\pi$, so that all plots seem to be $\pi/2$
periodic, and, in fact, symmetric upon reflection with respect to either
$\theta=0$ or $\theta$ $=\pi/2$ lines. \ We also consider $H,$ in the range
$0.1<H\lesssim0.45$ such that $g_{Q}\lesssim1$ (as explained before). \ It is
clear from Fig. 1 that the stability and instability domains are quite
different in classical and quantum plasmas. Evidently, the quantum parameter
$H$ shifts the MI domains in the $k\theta$-plane. It is, of course, hard to
find the common instability regions for both classical and quantum cases.
Increasing the value of $H$ $(\lesssim0.45)$ gives no effective change, except
in reducing the stable regions around $k\sim1.$

In particular, considering the modulation of the QIAWs along the carrier wave
vector $\mathbf{k}$ (i.e., the case of $\theta=0),$ the QIAW is shown to be
modulational unstable (see Fig. 2). This is in contrast to the classical case
where the IAWs are known to be stable under parallel modulation \cite{Kako}.
Physically, the QIAW under parallel modulation becomes unstable due to the
nonlinear self-interactions originating from the zeroth harmonic modes or slow
modes as well as the second harmonic modes (since $P$ and $Q$ are always
negative). However, the case of modulation perpendicular to the wave vector
$\mathbf{k}$ ($\theta=\pi/2$) gives rise stable wave propagation, an agreement
with classical plasmas \cite{Kako}. Here the wave is stable due to the second
harmonic self-generation of the modes, since $P>0$ and $Q<0.$ Thus, for the MI
to occur the second harmonic mode is very essential in order to maintain the
same sign of $P$ and $Q$.

On the other hand, the MI growth rate as given by Eq. (\ref{e19}) can be
calulated as depicted in Fig. 3. The upper panel (corresponding to QIAWs)
shows that the growth rate can \ be reduced by increasing the obliqueness of
modulation giving rise cut-offs at lower wave numbers of modulation. The
quantum parameter $H$ has no significant role in reducing such growth rate. In
the classical case, the maximum growth rate reduces slightly with a small
change of the angle of modulation (see lower panel of Fig. 3)


\begin{figure}[ht]
\includegraphics[width=0.9\columnwidth]{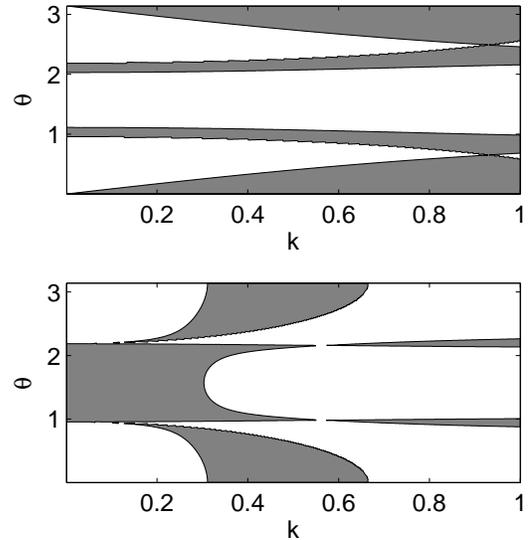}
\caption{The equation $PQ=0$ is contour plotted in the $k\theta$-plane to
show the stable and unstable regions in quantum (upper panel) and classical
(lower panel) plasmas [Eq.(\ref{e17})] for $H=0.3$. The white or stable (gray
or unstable) areas correspond to the regions in the $k\theta$-plane where
$PQ<0$ ($PQ>0$).}
\end{figure}

\begin{figure}[ht]
\includegraphics[width=0.9\columnwidth]{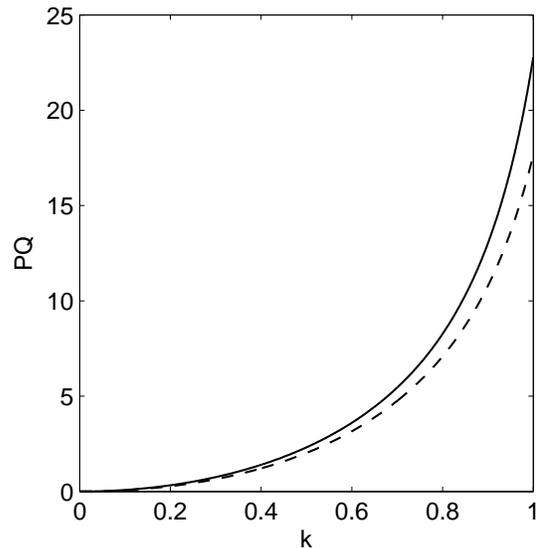}
\caption{The case of parallel modulation ($\theta=0$): $PQ$ is plotted against
$k$ \ to show that the QIAW is unstable ($PQ>0$). The solid and dashed lines
respectively correspond to the cases where $H=0.3$ and $H=0.45.$}
\end{figure}

\begin{figure}[ht]
\includegraphics[width=0.9\columnwidth]{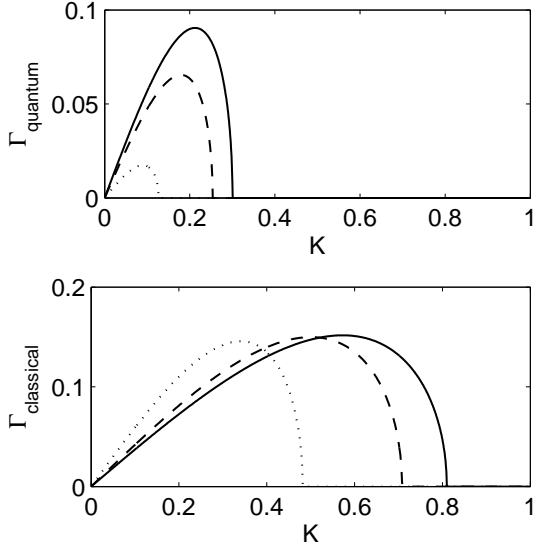}
\caption{The MI growth rate $\Gamma$ is shown\ with respect to the wave number
of modulation, $K$.\ \ \ The growth rate can be quenched by the obliqueness
parameter $\theta$ in quantum plasmas (upper panel). In the classical case
(lower panel), the maximum growth rate is reduced slightly with a small change
of $\theta$. The solid, dashed and dotted lines correspond to the
values\ $\theta=0.99,1.0,1.05$\ respectively.}
\end{figure}


\section{Variational principle and 2D-evolution}

The DS-like equations (\ref{15}) and (\ref{e16}) describes the case of general
2D-evolution of slowly varying weakly nonlinear ion-acoustic waves in the
quantum regime. Here we start by pointing out a variational principle for the
DS-like system. Introducing the Lagrangian density
\begin{align}
\mathcal{L}  &  =\frac{1}{2}\left(  \phi^{\ast}\frac{\partial\phi}%
{\partial\tau}-\phi\frac{\partial\phi^{\ast}}{\partial\tau}\right)  -P_{1}%
^{q}\left\vert \frac{\partial\phi}{\partial\xi}\right\vert ^{2}-P_{2}%
^{q}\left\vert \frac{\partial\phi}{\partial\eta}\right\vert ^{2}+\frac{1}%
{2}Q_{11}^{q}\left\vert \phi\right\vert ^{4}\nonumber\\
&  +\frac{Q_{22}^{q}}{S_{1}}\left[  \frac{1}{2}R_{1}\left(  \frac{\partial
^{2}u}{\partial\xi^{2}}\right)  ^{2}+\frac{1}{2}R_{2}\left(  \frac
{\partial^{2}u}{\partial\eta\partial\xi}\right)  ^{2}+S_{1}\frac{\partial
^{2}u}{\partial\xi^{2}}\left\vert \phi\right\vert ^{2}\right]  \label{e22}%
\end{align}
where the action functional is $\mathcal{A}(\phi,\phi^{\ast},u)=\int
\mathcal{L}d\xi d\tau,$ and $\partial^{2}u/\partial\xi^{2}\equiv\psi$ plays
the role of a potential. We obtain eqs. (\ref{15}) and (\ref{e16}) by varying
$\phi$, $\phi^{\ast}$ and $u$ and minimizing the action as usual. It is
straightforward to show that (\ref{15}) and (\ref{e16}) posses constants of
motion, representing the conservations of the number of high frequency quanta
\begin{equation}
\frac{d}{d\tau}\int_{-\infty}^{\infty}\int_{-\infty}^{\infty}\left\vert \phi\right\vert ^{2}d\eta
d\xi=0
\end{equation}
and of longitudinal momentum
\begin{equation}
\frac{d}{d\tau}\int_{-\infty}^{\infty}\int_{-\infty}^{\infty}\left[  \phi\frac{\partial\phi^{\ast}%
}{\partial\xi}-\phi^{\ast}\frac{\partial\phi}{\partial\xi}\right]  d\eta
d\xi=0
\end{equation}
as well as transverse momentum
\begin{equation}
\frac{d}{d\tau}\int_{-\infty}^{\infty}\int_{-\infty}^{\infty}\left[  \phi\frac{\partial\phi^{\ast}%
}{\partial\eta}-\phi^{\ast}\frac{\partial\phi}{\partial\eta}\right]  d\eta
d\xi=0 .
\end{equation}
Furthermore, since no explicit time dependence occurs in the Lagrangian, the
Hamiltonian $\int_{-\infty}^{\infty}\int_{-\infty}^{\infty}\mathcal{H}d\eta d\xi$, derived from $\mathcal{L}$ in (\ref{e22}), is also conserved.
The existence of a Variational principle makes it possible to use trial
functions as a means to obtain approximate solutions. A good review on this
topic for wave equations similar to (\ref{15}) and (\ref{e16}) is given by
Ref. \cite{Anderson-review}. A thorough variational study of (\ref{15}) and
(\ref{e16}) for the general case is beyond the scope of the present paper.
However, in order to illustrate the usefulness of the variational technique,
we will state a variational result that applies when a cylindrically symmetric
collapse is possible. For the general case, Eqs. (\ref{15}) and (\ref{e16}) do
not admit a cylindrically symmetrical collapse. However, for geometries and
parameter regimes where the coefficients fulfill either $R_{2}\ll R_{1}$, or
$Q_{22}^{q}S_{1}/R_{1}\ll Q_{11}^{q}$, our system reduces to a 2D NLS equation
for which cylindrically symmetrical solutions are possible. \ Assuming that at
least one of the above strong inequalities apply, and that we have
cylindrically symmetric initial conditions (to be defined below), the
evolution equation for our system can be written
\begin{equation}
i\frac{\partial\phi}{\partial\tau}+\frac{1}{r}\frac{\partial}{\partial
r}\left(  r\frac{\partial\phi}{\partial r^{2}}\right)  +Q_{\mathrm{eff}}%
|\phi|^{2}\phi=0 \label{e23}%
\end{equation}
where we have introduced the radial coordinate $r^{2}=\xi^{2}/P_{1}^{q}%
+\eta^{2}/P_{2}^{q}$ \cite{Cylindrical-note}, and assumed, cylindrical
symmetry, i.e. that the spatial dependence is only through this variable . The
coefficient $Q_{\mathrm{eff}}$ is either $Q_{11}^{q}$ (in case the reduction
from (\ref{15}) and (\ref{e16}) was due to the inequality $Q_{22}^{q}%
S_{1}/R_{1}\ll Q_{11}^{q}$) or $Q_{\mathrm{eff}}=Q_{11}^{q}+Q_{22}^{q}%
S_{1}/R_{1}$ (in case the reduction was due to $R_{2}\ll R_{1}$). For the case
of $Q_{\mathrm{eff}}>0$, collapse is possible. A good estimate for the
collapse threshold can be obtained by using the variational formulation, with
a trial function of the form $\phi(\tau,r)=F(\tau)\sec\mathrm{h}%
(r/f(\tau))\exp(ib(\tau)r^{2})$, see e.g. Ref. \cite{Collapse-ref} for
mathematical details. The result is that there will be a collapse towards a
zero radius of the pulse profile, $f(\tau)\rightarrow0$ within a finite time,
provided the initial profile is sufficiently intense and well localized, as
expressed by the collapse condition
\begin{equation}
Q_{\mathrm{eff}}f^{2}(\tau=0)\left\vert F^{2}(\tau=0)\right\vert \gtrsim1.35
\label{e24}%
\end{equation}
The accuracy of this collapse condition is well supported by numerical
calculations. Naturally, for cases where cylindrical symmetry does not apply,
the evolution will be much more complicated than indicated by our simple example.

\section{ Conclusion}

We have investigated the nonlinear propagation of QIAW packets in a 2D quantum
plasma. A multiple scale technique is used to derive a coupled set of
nonlinear partial differential equations, which governs the dynamics of
modulated QIAW packets. The set of equations, in particular (i.e., modulation
along the direction of the pump carrier wave), is shown to be reducible to a
well-known Davey-Stewartson (DS)-like equations \cite{Davey}. The latter can
be studied for the 2D wave collapse in dense plasmas. The oblique MI of QIAWs
is then studied by means of a corresponding NLS equation. It is found that the
quantum parameter $H$ has the significant role in shifting the instability
domains around the $k\theta$-plane. The case of parallel modulation, which is
known to be stable in classical plasmas \cite{Kako}, is shown to be unstable
in quantum plasmas. In contrast to the classical case, the MI growth rate in
quantum plasmas can significantly be reduced by increasing the angle of
modulation. 

It is to be mentioned that  the numerical simulation of the  DS equations
could furnish more convincing evidence for the process of soliton collapse,
where the quantum parameter $H$ may play a crucial role in accelerating or
decelarating the collapse process. However, it needs extra effort, and could
be an open issue to be explored in future studies. The integrability as well
as the dromion solution, if there be any, of the system will also be an
another investigation, but beyond the scope of the present work.

\acknowledgments

A. P. M. gratefully acknowledges support from the Kempe Foundations. This research is supported by the European Research Council under Contract No. 204059-QPQV and the Swedish Research Council under Contract No. 2007-4422



\section*{APPENDIX A}

The coefficients $P_{j}^{q},Q_{j}^{q},R_{j}^{q},S_{j}^{q}$ appearing in Eqs.
(\ref{e13}) and (\ref{e14}) are given as follows.
\begin{eqnarray*}
&&
P_{1,2}^{q}\equiv\frac{1}{2}\frac{\partial^{2}\omega}{\partial k_{x,y}^{2}%
}=\frac{\omega^{3}}{2k^{2}}\Bigg[
\frac{1-\omega^{2}}{\omega^{2}}+\frac{v_{g(x,y)}}{\omega^{3}}\left(
\frac{3k^{2}v_{g(x,y)}}{\omega}-4k_{x,y}\right)  
\nonumber \\ &&\qquad \qquad
+\frac{H^{2}}{4\Lambda^{2}}\left(  1-\frac{H^{2}k_{x,y}^{2}}{\Lambda}\right)
\Bigg]  ,
\\ &&
P_{3}^{q}=\frac{1}{2}\left[  \frac{v_{g}}{k}\left(  \frac{3kv_{g}}{\omega
}-4\right)  -\frac{\omega^{3}H^{4}}{4\Lambda^{3}}\right]  \sin2\theta,
\end{eqnarray*}
and
\begin{widetext}
\begin{align*}
Q_{1}^{q}  &  = -(k_{x}\lambda_{u}+k_{y}\lambda_{v})-\frac{1}{2}%
\omega\lambda_{i}-\frac{\omega^{3}}{k^{2}\Lambda^{2}}\left[
\frac{5}{2}\lambda_{e}\left(  \frac{1}{9}+\frac{H^{2}k^{2}}{4}\right) 
-\frac{1}{2\Lambda}\left(  \frac{7}{27}+\frac{3}{4}H^{2}k^{2}\right)
-\lambda_{\phi}%
\right]  \\
&  + \left\{  k_{x}A_{1}+\frac{\omega^{3}}{2k^{2}}\left(  \frac{3k_{y}^{2}%
}{\omega^{2}}+\frac{\Lambda_{0}}{\Lambda^{2}}\right)  \left[  \frac{k^{2}%
}{\omega^{2}}+\frac{1}{\Lambda^{2}}\left(  \frac{5}{9}-\frac{H^{2}k^{2}}%
{4}\right)  \right]  -\frac{\omega k_{y}^{2}}{\Lambda^{2}k^{2}}\right\}
.
\end{align*}
\end{widetext}
In the expression for $Q_{1}^{q}$ the first (second) term in the square
brackets is due to the first and second harmonic modes (the first and zeroth
harmonic modes).
\begin{align*}
Q_{2}^{q}  &  =\frac{\omega v_{gx}}{2}\left[  -3\left(  1+\frac{2k_{x}v_{gx}%
}{\omega}\right)  +\frac{\omega^{2}\tilde{\Lambda}}{\Lambda^{2}k^{2}}\right]
,\\
Q_{3}^{q}  &  =k_{x}\left(  1-3v_{gy}^{2}\right)  ,
\\
Q_{4}^{q}  &  =-k_{y}-3v_{gy}\left(  2v_{gx}k_{x}+\frac{\omega}{2}\right)
+\frac{\omega^{3}v_{gy}\tilde{\Lambda}}{2\Lambda^{2}k^{2}},\\
Q_{5}^{q}  &  =k_{x}k_{y}\left[  \frac{3}{2\omega}\left(  1+\frac{2v_{gx}%
k_{x}}{\omega}\right)  -\frac{\omega\tilde{\Lambda}}{2\Lambda^{2}k^{2}%
}\right]  ,
\\
Q_{6}^{q} & =k_{x}\left[  \frac{v_{gy}\Lambda_{0}}{\Lambda^{2}}+\frac{k_{y}%
}{\omega^{2}}\left(  \frac{2k^{2}}{\omega}+3k_{y}v_{gy}\right)  \right]  ,
\\
R_{1}^{q} & =v_{gx}\left(  3v_{gx}^{2}-1\right)  ,R_{2}^{q}=v_{gx}\left(
9v_{gy}^{2}-1\right)  ,
\\
R_{3}^{q} & =v_{gy}\left(  9v_{gx}^{2}-1\right)  ,R_{4}^{q}=v_{gy}\left(
3v_{gy}^{2}-1\right)  ,
\\
S_{1}^{q} & =\frac{1}{\omega^{2}}\left[  v_{gx}\left(  A_{1}\omega^{2}%
+3k_{x}k_{y}v_{gy}\right)  +k_{x}^{2}-k_{y}^{2}\right]  ,
\\
S_{2}^{q} & =\frac{1}{\omega^{2}}\left[  v_{gy}\left(  A_{1}\omega^{2}+3k_{y}%
^{2}v_{gx}+v_{gx}\Lambda_{0}\frac{\omega^{2}}{\Lambda^{2}}\right)
\right. 
\\ &\qquad\qquad \left.
+k_{y}\left(  k_{x}+\frac{2v_{gx}k^{2}}{\omega^{2}}\right)  \right]  ,
\\
S_{3}^{q} & =\frac{k_{x}k_{y}}{\omega^{2}}\left(  3v_{gx}^{2}-1\right)
,S_{4}^{q}=A_{2}v_{gy},
\end{align*}
where
\begin{align*}
\tilde{\Lambda}  &  =\frac{1}{3}+\frac{3}{4}H^{2}k^{2},\Lambda_{0}=\frac{5}%
{9}+\frac{H^{2}k^{2}}{4},\\
\bar{\Lambda}  &  =\frac{1}{3}+H^{2}k^{2},
\\ 
\lambda_{\phi}  &  =\frac{\omega^{4}(4+27H^{2}k^{2})-72\bar{\Lambda}%
\Lambda^{2}k^{4}}{72\omega^{2}\Lambda^{2}\left[  k^{2}\bar{\Lambda}%
^{2}(1-4\omega^{2})-\omega^{2}\right]  },
\\
\lambda_{e}  &  =\frac{k^{2}}{\omega^{4}}\left[  k^{2}+\lambda_{\phi}%
\omega^{2}(1-4\omega^{2})\right]  ,
\\
\lambda_{i}  &  =\frac{k^{2}}{\omega^{4}}\left(  k^{2}+\lambda_{\phi}%
\omega^{2}\right)  ,
\\
\lambda_{u,v}  &  =\frac{k_{x,y}}{\omega}\left(  \lambda_{\phi}+\frac{k^{2}%
}{2\omega^{2}}\right)  ,
\\
A_{1} & =v_{gx}\frac{\Lambda_{0}}{\Lambda^{2}}+\frac{2k^{2}k_{x}}{\omega^{3}%
}+\frac{3k_{y}}{\omega^{2}}\left(  v_{gx}k_{y}+v_{gy}k_{x}\right)  ,
\\
A_{2} & =v_{gy}\frac{\Lambda_{0}}{\Lambda^{2}}+\frac{k_{y}}{\omega^{3}}\left(
2k^{2}+3\omega k_{y}v_{gy}\right) .
\end{align*}


\section*{APPENDIX B}

The coefficients corresponding to classical electron-ion plasmas can be
obtained by considering a 2D fluid model with contininuity and momentum
equations for cold ions and Boltzmann distributed electrons [see, e.g., the
Eqs. (5)-(8) in Ref. \cite{Kako}]. The normalizations and the corresponding
fluid equations for the variables can be recovered by simply replacing the
Fermi temperature $T_{Fe}$ by the classical temperature $T_{e}$ with
considering the electron pressure as $p_{e}=k_{B}T_{e}n_{e}$ instead of the
Fermi pressure, and disregarding the Bohm potential term proportional to
$\hbar$. These coefficients are presented for a general interest of the
readers to study the dynamics of ion-acoustic waves from the evolution
equations of the forms (\ref{e13}), (\ref{e14}) or (\ref{15}), (\ref{e16}) in
2D classical electron-ion plasmas.

Thus, the coefficients $P_{j}^{c},Q_{j}^{c},R_{j}^{c},S_{j}^{c}$ to be
appeared in equations like (\ref{e13}) and (\ref{e14}) in the case of
classical plasmas can be given as follows:
\begin{align*}
P_{1,2}^{c} & \equiv\frac{1}{2}\frac{\partial^{2}\omega}{\partial k_{x,y}^{2}%
}=\frac{\omega^{3}}{2k^{2}}\left[  \frac{1-\omega^{2}}{\omega^{2}}%
+\frac{v_{g(x,y)}}{\omega^{3}}\left(  \frac{k^{2}v_{g(x,y)}}{\omega}%
-4k_{x,y}\right)  \right]  ,
\\
P_{3}^{c} & =\frac{v_{g}}{2k}\left(  \frac{kv_{g}}{\omega}-4\right)  \sin2\theta,
\\
Q_{1}^{c}  &  = -(k_{x}\lambda_{u}+k_{y}\lambda_{v})-\frac{1}{2}%
\omega\lambda_{i}+\lambda_{e}-\frac{k_{y}^{2}}{\omega^{2}} \\
& + k_{x}A_{1}+\frac{\omega}{2}\left(  \frac{k_{y}^{2}}{\omega^{2}%
}-1\right)  ,
\end{align*}
where, in the expression for $Q_{1}^{c}$, the first (second) term in the square
brackets is due to the first and second harmonic modes (the first and zeroth
harmonic modes). Moreover, 
\begin{align*}
Q_{2}^{c}  &  =\frac{\omega v_{gx}}{2}\left[  -\left(  1+\frac{2k_{x}v_{gx}%
}{\omega}\right)  +\frac{\omega^{2}}{k^{2}}\right]  ,
\\
Q_{3}^{c}  &  =k_{x}\left(  1-v_{gy}^{2}\right)  ,
\\
Q_{4}^{c}  &  =-k_{y}-v_{gy}\left(  2v_{gx}k_{x}+\frac{\omega}{2}\right)
+\frac{\omega^{3}v_{gy}}{2k^{2}},
\\
Q_{5}^{c}  &  =k_{x}k_{y}\left[  \frac{3}{2\omega}\left(  1+\frac{2v_{gx}%
k_{x}}{\omega}\right)  -\frac{\omega}{2k^{2}}\right]  ,
\\
Q_{6}^{c} & =k_{x}\left[  -v_{gy}+\frac{k_{y}}{\omega^{2}}\left(  \frac{2k^{2}%
}{\omega}+k_{y}v_{gy}\right)  \right]  ,
\end{align*}
\begin{align*}
R_{1}^{c} & =v_{gx}\left(  v_{gx}^{2}-1\right)  ,
\\
R_{2}^{c} & =v_{gx}\left(3v_{gy}^{2}-1\right)  ,
\\
R_{3}^{c} & =v_{gy}\left(  3v_{gx}^{2}-1\right)  ,R_{4}^{c}=v_{gy}\left(
v_{gy}^{2}-1\right)  ,
\end{align*}
and
\begin{align*}
S_{1}^{c} & =\frac{1}{\omega^{2}}\left[  v_{gx}\left(  A_{1}\omega^{2}+k_{x}%
k_{y}v_{gy}\right)  +k_{x}^{2}-k_{y}^{2}\right]  ,
\\
S_{2}^{c} & =\frac{1}{\omega^{2}}\bigg[
v_{gy}\left(  A_{1}\omega^{2}+k_{y}^{2}v_{gx}-v_{gx}\omega^{2}\right) 
\\ &\qquad
	+k_{y}\left(  k_{x}+\frac{2v_{gx}k^{2}}{\omega^{2}}\right)
\bigg]  ,
\\
S_{3}^{c} & =\frac{k_{x}k_{y}}{\omega^{2}}\left(  v_{gx}^{2}-1\right)  ,S_{4}%
^{q}=A_{2}v_{gy},
\end{align*}
where
\begin{align*}
\lambda_{\phi}  &  =\frac{\omega^{4}-2k^{4}}{2\omega^{2}\left[  k^{2}%
(1-4\omega^{2})-\omega^{2}\right]  },\\
\lambda_{e}  &  =\frac{k^{2}}{\omega^{4}}\left[  k^{2}+\lambda_{\phi}%
\omega^{2}(1-4\omega^{2})\right]  ,
\\
\lambda_{i} & =\frac{k^{2}}{\omega^{4}}\left(  k^{2}+\lambda_{\phi}\omega
^{2}\right)  ,\lambda_{u,v}=\frac{k_{x,y}}{\omega}\left(  \lambda_{\phi}%
+\frac{k^{2}}{2\omega^{2}}\right)  ,
\\
A_{1} & =-v_{gx}+\frac{2k^{2}k_{x}}{\omega^{3}}+\frac{k_{y}}{\omega^{2}}\left(
v_{gx}k_{y}+v_{gy}k_{x}\right)  ,
\\
A_{2} & =-v_{gy}+\frac{k_{y}}{\omega^{3}}\left(  2k^{2}+\omega k_{y}%
v_{gy}\right)
\end{align*}
where $\omega^{2}=k^{2}/(1+k^{2})$ is the normalized classical dispersion
relation ($\omega \rightarrow \omega/\omega_{pi} $ and $k \rightarrow kc_s/\omega_{pi}$). We note that the difference in the dispersion relation as compared to Eq. (10) stems from the use of a classical (i.e. not a Fermi) equation of state. 

\end{document}